\documentclass[a4paper]{article}

\usepackage[table,xcdraw]{xcolor}
\usepackage{enumitem}
\usepackage{spconf,graphicx}
\usepackage{tikz,tikz-qtree}
\usepackage{booktabs}
\usepackage{multirow}
\usepackage{INTERSPEECH2022}

\title{Analyzing the impact of SARS-CoV-2 variants on respiratory sound signals}
\name{Debarpan Bhattacharya$^1$,
Debottam Dutta$^1$, Neeraj Kumar Sharma$^2$, Srikanth Raj Chetupalli$^2$, Pravin Mote$^1$, Sriram Ganapathy$^1$, Chandrakiran C$^3$, Sahiti Nori$^3$, Suhail K K$^3$, Sadhana Gonuguntla$^4$, Murali Alagesan$^5$}
\address{
  $^1$LEAP lab, Indian Institute of Science, Bangalore, India,
$^2$Erlangen, Germany,
  $^3$Ramaiah Medical College Hospital, Bangalore, India,
  $^4$General Hospital, Hoskote, Bangalore, India,
  $^5$PSG Institute of Medical Sciences and Research, India
  }
\email{sriramg@iisc.ac.in}
\ninept 
\begin{document}
\maketitle
\begin{abstract}
The COVID-19 outbreak resulted in multiple waves of infections that have been associated with different SARS-CoV-2 variants. Studies have reported differential impact of the variants on respiratory health of patients. We explore whether acoustic signals, collected from COVID-19 subjects, show computationally distinguishable acoustic patterns suggesting a possibility to predict the underlying virus variant. We analyze the Coswara dataset which is collected from three subject pools, namely, i) healthy, ii) COVID-19 subjects recorded during the delta variant dominant period, and iii) data from COVID-19 subjects recorded during the omicron surge. Our findings suggest that multiple sound categories, such as cough, breathing, and speech, indicate significant acoustic feature differences when comparing COVID-19 subjects with omicron and delta variants. The classification areas-under-the-curve are significantly above chance for differentiating subjects infected by omicron from those infected by delta. Using a score fusion from multiple sound categories, we obtained an area-under-the-curve of $89\%$ and $52.4\%$ sensitivity at $95\%$ specificity. Additionally, a hierarchical three class approach was used to classify the acoustic data into healthy and COVID-19 positive, and further COVID-19 subjects into delta and omicron variants providing high level of 3-class classification accuracy. These results suggest new ways for designing sound based COVID-19 diagnosis approaches.
%   On account of massive disruption in human life worldwide, owing to COVID-19 outbreak, there have been numerous efforts put together by researchers to mitigate the vulnerability of the pandemic. Rapid diagnosis of COVID-19 on a massive scale has been instrumental to achieve the goal. Respiratory acoustics based diagnosis of COVID-19 gathered significant attraction from the researchers. The growing interest along that direction gave birth to large acoustic datasets from multiple countries of the world. For example, Coswara, based in India and being one of the major public datasets, contains 681 COVID-19 positive samples. With the help of those datasets, few acoustic analysis based studies of positive and healthy samples have been reported in the literature recently. But, there has been no acoustic analysis which intends to study the effect of different variants of COVID-19 on respiratory signals, up to the best of our knowledge. Such a study is carried out based on Coswara dataset and reported in this paper. While it validates few important claims regarding the COVID variants, put forward by the medical community, on the other hand, it indicates interesting findings on the effect of variants on respiratory acoustics.
\end{abstract}
\noindent\textbf{Index Terms}: COVID-19, SARS-CoV-2 variants, Omicron, cough, breathing, vowel, counting, and speech.

\section{Introduction}
Since the global outbreak of COVID-19 in March 2020, the world has been responding to disruptions across the health, education, and economic sectors. Among various measures to contain the pandemic, rapid diagnosis and isolation of COVID-19 patients helped to control the scale of SARS-CoV-2 infections \cite{pramesh2021choosing}.
Over the time scale of more than 24 months, the SARS-CoV-2 virus has continued to make mutations in its gene structure. Studies such as He et. al. \cite{he2021sars} suggest the possibility of numerous mutations in future, and highlight the need for adopting diagnosis and treatment strategies to combat them in the shortest possible time. Such mutated variants of the virus are identified as variants of concern (VoC).
\begin{figure*}[t]
    \centering
    \input{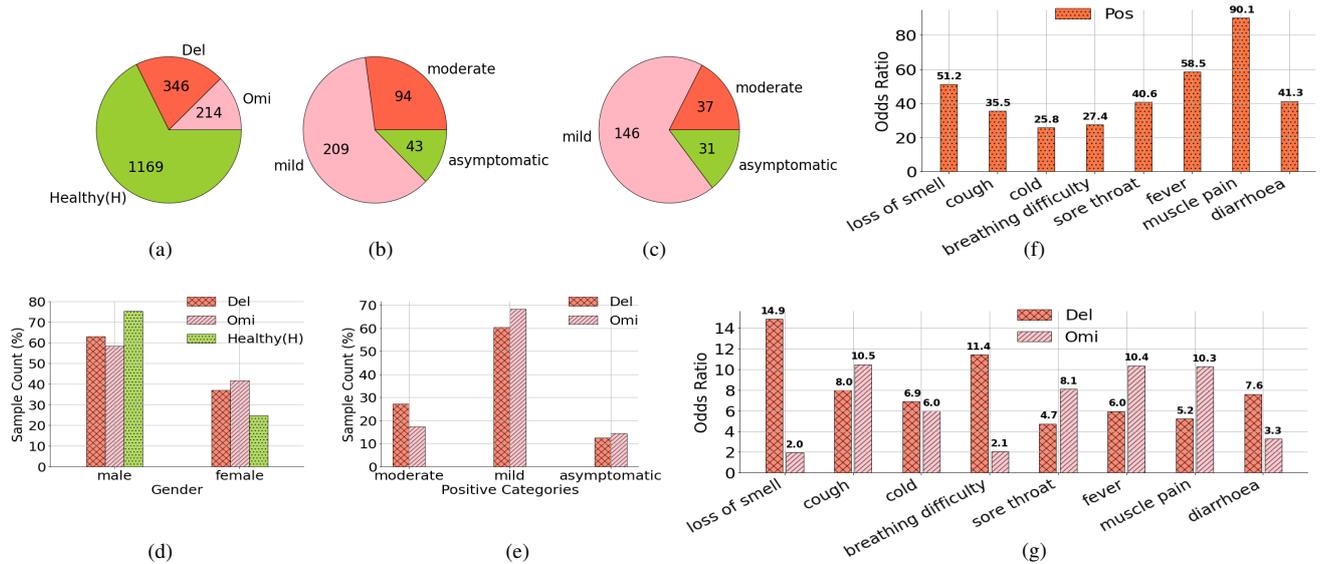}
    \vspace{-0.1in}
    \caption{(a) Pie plot representing number of samples for different subject categories as in Table \ref{tab:subject_sound_categroies}; (b),(c) pie plot representing number of samples for different COVID severity conditions for delta and omicron positive samples respectively; (d) gender metadata plot; (e) COVID severity plot for positive samples in percentage; (f) odds ratio of different symptoms for positive samples; (g) comparison of odds ratio for different symptoms for omicron and delta positive samples.}
    \label{fig:dev_set_metadata}
\vspace{-0.15in}
\end{figure*}

% of different variants has also hampered the efficacy of the vaccines deployed across the world
% \cite{pilishvili2021effectiveness,juno2021boosting} as well as antibody based treatments \cite{kozlov2021omicron}. 
Two such VoCs are the delta (B.1.617.1) and the omicron (B.1.1.529) variants, which have caused significant rise in the COVID-19 cases in the Indian subcontinent \cite{mohapatra2022twin} in two non-overlapping time spans. The delta variant related COVID-19 outbreak happened in April-June 2021 and the omicron  variant related outbreak happened in Jan-Feb 2022 \cite{gopal2022analysis, ranjan2022omicron}. Interestingly, a recent study by Hui et. al. \cite{hui2022sars} showed that omicron variants had a higher replication rate in bronchus and lesser replication rate in the lungs, as compared to the delta variant. 
% Sigel et. al. \cite{sigal2022milder} argue that the combination of vaccine induced immunity in the population and the viral properties of the omicron to be the reason behind the reduced fatality and lung impact of this newer variant.
Drawn by this, we hypothesize that the respiratory sounds of COVID-19 patients might encode a variant dependent attribute.
That is, we are interested in exploring if the respiratory sound samples of a COVID-19 patient can help identify the category of the variant causing the infection. To validate this hypothesis, this paper presents a diverse sets of experiments with sound samples collected from healthy and COVID-19 patients across multiple waves of SARS-CoV-2 infection in the Indian subcontinent.

% This question is also motivated by the ability of acoustic signals to differentiate lung related impairments versus upper respiratory tract related ailments.
To our knowledge, none of the COVID-19 audio datasets have recorded the SARS-CoV-2 variant information for the COVID-19 patients.
However, the Coswara dataset~\cite{sharma2020coswara}, an open-access COVID-19 audio dataset, provides the timestamp corresponding to the date each COVID-19 patient contributed their audio data to the dataset. As a work around to obtain information related to variant identity, we match the variant identity to the one which was widespread in the locality on the date of audio data contribution. Based on this, a majority of the COVID-19 patients contributed data to the Coswara dataset during the outbreak of the delta and omicron variants in India.
The Coswara dataset contains audio signals from nine sound categories, collected via crowdsourcing and majorly from India.
% Although it is reasonable to have label noise while considering all positives in second wave to be originating from the delta variant and all positives in third wave to be from the omicron variant, the noise must be minimal given the dominance of the variants during respective timelines~\cite{ranjan2022omicron}.
%%%%%%%%%%%%%%%%%%
% State the contributions clearly below in one paragraph
% %%%%%%%%%%%%%%%%
This study aims to classify delta and omicron variants based on the claim, set forth by the medical community, that they are lower and upper respiratory diseases respectively. It presents discriminability analysis, both statistically and using LSTM classifiers, with the 9 types of respiratory sounds present in the Coswara dataset.

We observe that multiple sound categories, such as cough, breathing, and speech, indicate significant acoustic feature difference when comparing COVID-19 subjects with omicron variant versus delta variant.
Further, the classification AUCs are significantly above chance for a omicron versus delta variant detection using sound samples only.
Using fusion strategy on predictions obtained from multiple sound categories, we obtain AUC of $89\%$ and sensitivity of $52.4\%$ at $95\%$ specificity.
%A highly diagonally dominant confusion matrix is obtained in this case.
% \cite{janeiro2021can} shows a study where the authors correlated the waves of different SARS-CoV-2 variants in the region of Galicia, Spain with the number of cycles to determine samples to be positive in RT-PCR test. \cite{lebatteux2022machine} proposed machine learning based algorithm KEVOLVE which is able to identify genomic signatures of SARS-CoV-2 variants. But, to the best of our knowledge, this is the first study of its kind which attempts to differentiate different variants based on the acoustic signal properties.
% \vspace*{100px}
% \begin{table}[]
% \centering
% \rowcolors{2}{white}{gray!10}
% \resizebox{\columnwidth}{!}
% {
% \begin{tabular}{@{}lll@{}}
% \toprule
% \textbf{Sound Category} & \multicolumn{1}{c}{\textbf{}} & \multicolumn{1}{c}{\textbf{Description}} \\ \midrule
% Breathing-shallow/deep &  & \begin{tabular}[c]{@{}l@{}}Few cycles of inspiration and expiration:\\ without (shallow)/with (deep) exertion on the lungs\end{tabular} \\
% Cough-shallow/heavy &  & \begin{tabular}[c]{@{}l@{}}Few bouts of cough:\\ without (shallow)/with (heavy) exertion on the lungs\end{tabular} \\
% Vowel-[u]/[i]/[\ae] &  & \begin{tabular}[c]{@{}l@{}}Sustained phonation of three different vowels:\\ $[$u] as in boot, [i] as in beet, and [\ae] as in bat\end{tabular} \\
% Counting-normal/fast &  & \begin{tabular}[c]{@{}l@{}}Counting numbers from 1 to 20 in:\\ normal/fast pace\end{tabular} \\ \bottomrule
% \end{tabular}
% }
% \caption{}
% \label{tab:my-table}
% \end{table}

\begin{table}[]
\centering
\rowcolors{2}{white}{gray!10}
% \resizebox{\columnwidth}{2.75cm}
{
\begin{tabular}{@{}ll@{}}
\toprule
\textbf{Subject Category} &   \multicolumn{1}{c}{\textbf{Description}} \\ \midrule
Healthy ($H$) &   \begin{tabular}[c]{@{}l@{}}Healthy subjects, not infected with\\ SARS-CoV-2\end{tabular}  \\
Delta ($Del$) &   \begin{tabular}[c]{@{}l@{}}Infected with SARS-CoV-2 with\\ a strain earlier to Omicron outbreak\end{tabular} \\
Omicron ($Omi$) &  \begin{tabular}[c]{@{}l@{}}Infected with SARS-CoV-2 with\\ a strain during Omicron outbreak\end{tabular} \\
Positive ($Pos$) &  \begin{tabular}[c]{@{}l@{}}Del and Omi subjects pooled\end{tabular} \\\midrule
\multicolumn{1}{c}{\textbf{Sound Category}} &  \multicolumn{1}{c}{\textbf{Description}} \\ \midrule
Breathing-shallow/deep &  \begin{tabular}[c]{@{}l@{}}Few cycles of inspiration/expiration:\\ with shallow/deep exertion on lungs\end{tabular} \\
Cough-shallow/heavy &   \begin{tabular}[c]{@{}l@{}}Few bouts of cough:\\ with shallow/heavy exertion on   lungs\end{tabular} \\
Vowel-[u]/[i]/[\ae] &   \begin{tabular}[c]{@{}l@{}}Sustained phonation of three vowels:\\ $[$u] in boot, [i] in beet, [\ae] in bat\end{tabular} \\
Counting-normal/fast &  \begin{tabular}[c]{@{}l@{}}Counting numbers from $1$ to $20$ at:\\ normal/fast pace\end{tabular} \\ \bottomrule
\end{tabular}
}
%\vspace{0.1in}
\caption{Subject and sound categories analyzed in this study}
\vspace{-0.35in} 
\label{tab:subject_sound_categroies}
\end{table}

\section{Materials}
% As of February 2022, the Coswara dataset has $681$ positive and $1433$ negative samples. The database contains samples collected starting from May-2020 onwards. It also features the most diverse acoustic sound sample collection with $9$ categories of respiratory sounds recorded from the subjects.
% The set of samples in the dataset are grouped into 3 categories, namely, Omicron positive (251 samples), Delta positive (430 samples), healthy (1433 samples). Positive category means combination of Omicron positives and Delta positives. The samples, for which all the $9$ categories of sounds are not of good quality, are filtered out.
A subset of Coswara dataset based on selecting subjects with: all audio files being of good quality, the country location being India, age between 15-90 years, and belonging to either healthy or COVID-19 positive category, is drawn. This results in a dataset consisting of $1169$ healthy, and $560$ COVID-19 positive subjects. Based on the timestamp associated with the date of audio data contribution to the dataset, the $560$ COVID-19 subjects were further divided into two categories, namely, delta ($346$) and omicron (214).
Each subject contributed nine audio signals to this dataset. The sound sample categories are described in Table \ref{tab:subject_sound_categroies}.
In the past, several studies have shown evidence for diagnosing a multitude of pulmonary diseases via computational analysis of these sound categories, example see \cite{tinkelman1991analysis, al2012asthma, nathan2019assessment}. Apart from the sound recordings, a diverse set of metadata information was also collected consisting of age/gender, broad geographic location, current health status w.r.t. symptoms associated with COVID-19, as well as the COVID-19 testing status. In addition, the COVID-19 positive samples are classified as one of the categories: asymptomatic, mild, and moderate based on the severity of the symptoms in the subject. The samples were not collected from severe patients considering their health condition. 
The metadata statistics for the subset of the data used in this study in terms of gender, COVID-19 severity  and odds-ratio of symptoms, are reported in Figure~\ref{fig:dev_set_metadata}.
The odds-ratio is defined as the ratio of the subjects exhibiting a particular symptom and being positive with particular variant to the ratio
of subjects who have the same symptoms but are not positive with the particular variant. Odds ratio of more than 1 denotes positive correlation between the symptom and the disease.
% Let, X and Y are two events. The odds ratio of X and Y is defined as the ratio of odds of occurrence of X in presence of Y to odds of occurrence of X in absence of Y. Odds of occurrence of X can be calculated as ratio of occurrence of X to non-occurrence of X.
% It is a symmetric metric which means that X and Y can be interchanged in the definition. Odds ratio of more than 1 indicates positive correlation, value of less than 1 indicates negative correlation and value equal to 1 indicates no correlation among events X and Y.
As seen in figure \ref{fig:dev_set_metadata}, the odds-ratio of positive subjects is highest for the condition of muscle pain. Further, among the variants, delta variant had higher odds-ratio for 'loss-of-smell', 'breathing-difficulty' and 'diarrhea', while omicron variant had higher odds-ratio for 'cough', 'sore-throat', 'fever' and 'muscle-pain'.

\section{Methodology}
%\subsection{Pre-processing}

%Also, a buffer duration of $50$ ms is used to avoid discarding any samples on either side of a sample above the threshold.

% The audio sample was normalized to lie between $\pm~1$. This was followed by discarding low activity regions from the signal. Using a sound sample activity detection threshold of $0.01$, and a buffer size of $50~$msec on either side of a sample, any audio region with sample values greater than the threshold was retained.

\subsection{Feature extraction}
\label{sec:acoustic_features}
All the audio files are passed through a sound activity detector (SAD) after  normalizing the sample range to $\pm~1$. Subsequently, any audio sample residing below a threshold of $0.01$ is discarded. 
The log mel-spectrogram features are used for the analysis in the paper. Majority of the audio recordings have sampling rate of $48$kHz. All the audio recordings are first re-sampled to $44.1$ kHz. The log mel-spectrogram features are computed using a window of $1024$ samples, a stride of $441$ samples and with $64$ mel-filters. 
The delta and delta-delta features, which are first and second derivatives of mel-spectrogram along the time axis respectively, are appended to the mel-spectrogram. 
Thus, the features are of size $192\times N_k$,  where $N_k$ denotes number of short-time features obtained from k-th audio file.

% The log mel-spectrogram features were extracted using short-time windowed segments of size $1024$ samples ($23.2~$msec) and temporal hop of $441$ samples ($10$~ms), and a $64$ mel-filter filterbank. This resulted in a $64\times N_k$ dimensional feature matrix for the $k^{th}$ sound file, where $N_k$ represents the number of short-time frames.
% The mel-spectrogram features were appended with the first and second order temporal derivatives. The resulting $192\times N_k$ dimensional features were file-level mean and variance normalized. For brevity, we will refer to these features by \textit{mel-spec}.

\subsection{Statistical analysis}
\label{sec:statistical_analysis}
We pursue a statistical analysis comparing the acoustic features derived from subjects belonging to a pair of conditions. For this analysis, the mel-spec feature matrix computed for each audio file is averaged across the $N_k$ frames to obtain a $192\times1$ dimensional average mel-spec feature vector.  Subsequently, we obtain two populations of such feature vectors, one corresponding to each subject category. 
We do a dimension-wise (across $192$ dimensions) Mann-Whitney U test \cite{mann1947test} to statistically compare the feature values between two populations. The Mann-Whitney U test is a non-parametric test of the null hypothesis that, for randomly selected values of $X$ and $Y$ from two populations, the probability of $X$ being greater than $Y$ is equal to the probability of $Y$ being greater than $X$. 
For summarizing the statistical significance across all the feature dimensions, we compute the harmonic mean of p-values (HMP) from the $192$ p-values computed over each dimension of the average mel-spec feature vector. The HMP has been shown to be a robust measure for p-value summarization in high dimensional data analysis \cite{Wilson1195}. 
%We carry out this statistical analysis for each sound category.

\subsection{Classification}
Using the acoustic features computed in Section~\ref{sec:acoustic_features},  we pursue the  classifications tasks of  i) $Omi$-vs-$Del$, ii) $Omi$-vs-$H$, iii) $Del$-vs-$H$,  iv) $Pos$-vs-$H$, and  v) $Omi$-vs-$Del$-vs-$H$. Here, $Omi$ denotes the positive samples from the omicron variant, $Del$ denotes the positive samples from the delta variant and $H$ denotes the healthy subject samples.

For the three class hierarchical classification problem among $Omi$, $Del$ and $H$ samples, the given audio sample is first processed with the $Pos$-vs-$H$ classifier. If the sample is classified as positive, it is processed through the $Omi$-vs-$Del$ classifier.  

\subsection {Dataset division}

We divide the pool of four subject categories, $Omi$, $Del$, $Pos$ and $H$, into train, validation and test sets by drawing a random subset of $65\%$ of the subjects for train, $15\%$  for validation, and $20\%$ subjects for testing. The split of the omicron and delta positive samples is done in such a way that COVID-19  severity groups (asymptomatic, mild and moderate) are stratified.  %The group, named $Pos$, is formed by combining $Omi$ and $Del$  samples, as mentioned in Table \ref{tab:subject_sound_categroies}. 
%The train, val and test splits of positive class are the combinations of train, val and test splits of Omicron and Delta positives respectively. Splitting of the healthy category is done using the same train:val:test ratio.
For each of the classification tasks, the train and val splits are randomly sampled using $10$ different seeds, without changing the test split.  
%The $10$ fold validation results are averaged to increase the confidence about the classifier performance.
\subsection{Classifier model} 
% While the logistic regression and random forest based simple classifiers gave ~60\% AUC to classify COVID positives and negatives \cite{sharma2021second}, a strong baseline of Bi-LSTM based deep model performed much better to offer ~80\% AUC on Second DiCOVA challenge test set\cite{sharma2021second}.
We use the bidirectional long-short term memory (BLSTM)  based neural network architecture as the classification model. This model   served as a baseline in the Second DiCOVA Challenge~\cite{sharma2021second}. The architecture is composed of two BLSTM layers having 128 cells, followed by a feed-forward network having 64 neurons and a $\tanh(\cdot)$ non-linearity. The final output is a single neuron having sigmoid activation and the output represents the probability of positive class.
% The initial experimentation with the  traditional classifiers such as logistic regression and random forest gave a performance in the range of $60-70\%$ AUC on the validation folds for all the tracks. With the aim to provide a competent baseline, we opted using a deep learning framework with a cascade of two bi-directional long-short term memory (BiLSTM) and a fully connected layer. The initial BiLSTM layers with $128$ units were used to model the long-term dependencies in the audio signal. The output of the BiLSTM layers are of dimension $256\times T$. This is fed to a pooling layer which performs averaging along the time dimension to generate a sequence level embedding of $256\times1$. This output is fed to  a fully connected feedforward layer of $64$ nodes and a $\tanh(\cdot)$ non-linearity. The final layer resembles a two class logistic regression model with $64$ input nodes.
\\
\noindent {\bf Training:} 
The model is trained using audio segments, as described in \cite{sharma2021second}. Each log mel-spectogram is chunked into segments having contiguous frames, using $51$ frames window size with a stride of $10$ frames. All the segments generated from an audio recording are assigned the same class. During training, the minority class segments are over-sampled to reduce the impact of the class imbalance. The model is trained using the binary cross entropy (BCE) loss.
% For training the classifier, contiguous segments were extracted, with a $10$ frame stride, from the features matrix to obtain $192\times T$ fixed dimensional feature  representations. We choose $T$ as $51$ in the baseline system. The label of each chunk is the same as that of the audio file. Each mini-batch is composed of $1024$ feature matrices of size $192\times T$, randomly sampled from different audio files such that the proportion of COVID and Non-COVID labels is balanced. This oversampling of the  minority class is done to overcome the limitation of class imbalance. The binary cross entropy (BCE) loss, Adam optimizer with an initial learning rate of $0.0001,$ and $\ell_2$ regularization set to $0.0001$, were used to train the classifier. The learning rate was reduced by a factor of $10$ for a patience parameter set to three epochs. A dropout factor of $0.1$ was applied to the outputs of the first BiLSTM layer and the feedforward layer.
\\
\noindent {\bf Inference:} 
The test audio file  is chunked into multiple segments of $51$ frames. The output probabilities are calculated for all segments and their mean is computed to obtain the score for the audio file.
% Given an audio recording, $192\times T$ mel-spectrogram feature matrices (with a stride of $10$ frames) were extracted (similar to the training stage). These were input to the trained classifier and the output probability scores were obtained for each chunk. The average of the  probability scores from each segment was output as the COVID probability score of the audio file. On the blind test set, for each sound category, we draw inference by averaging the score obtained using a model trained on each validation fold.
\\
\noindent {\bf Performance Evaluation:}
We use the area-under-the-curve (AUC) measure of the receiver operating characteristic curve (ROC) \cite{FAWCETT2006861} for quantifying binary classifier performance. The AUC is computed using the trapezoidal rule \cite{scikit-learn}. An AUC of $0.5$ and $1$ indicate the  chance and best performance, respectively.

% \noindent {\bf Fusion:}
% Classifiers are trained separately for all the $9$ sound categories. A fusion probability score for an audio is computed by taking arithmetic mean of probability scores from the nine models.

\begin{figure}[t]
    \centering
    \includegraphics[width=8.5cm, height=7.5cm]{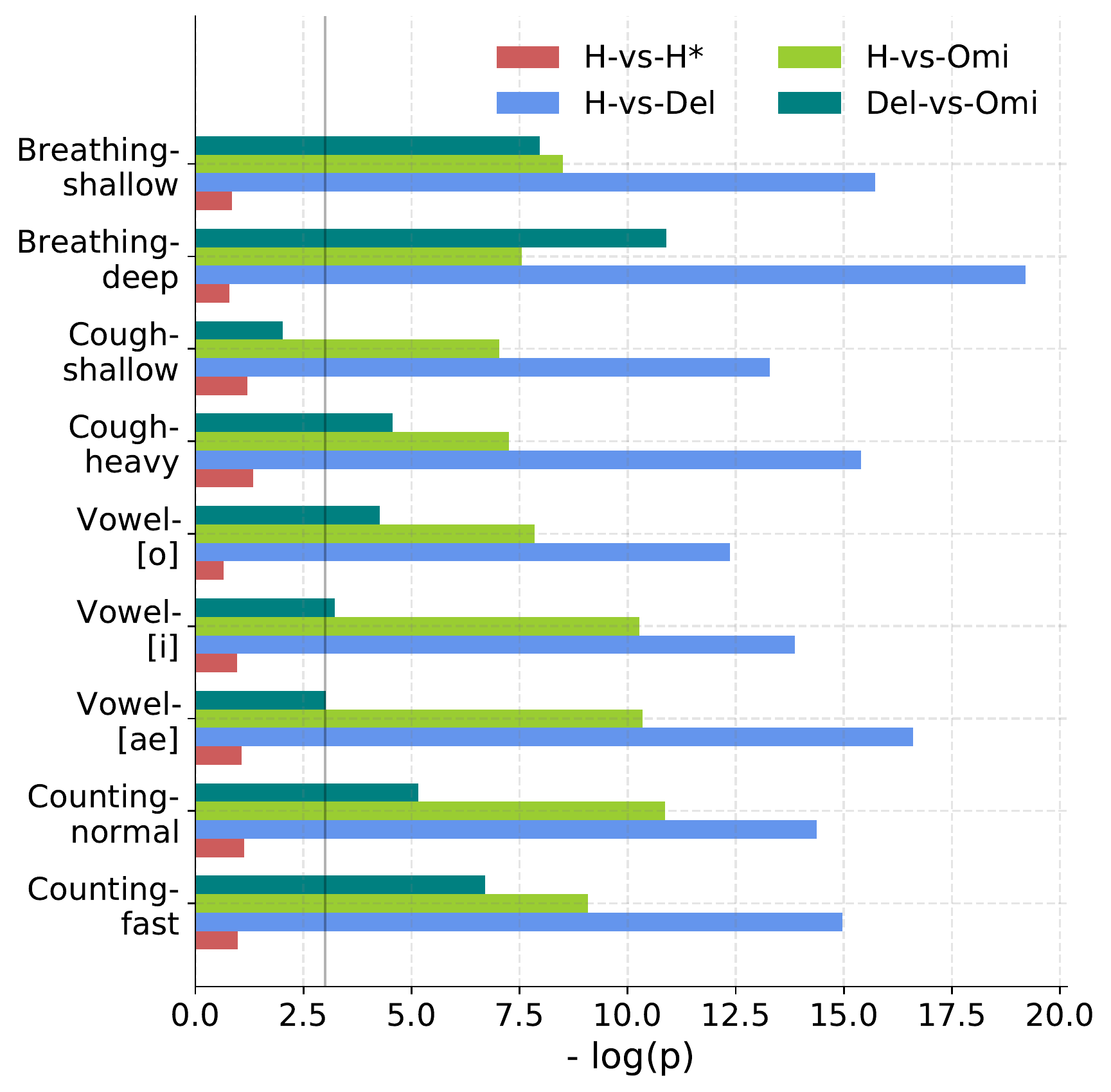}
    \caption{Illustration of statistical analysis using Mann-Whitney U test comparing acoustic features of different subject populations. The (gray) line indicates statistical significance corresponding to $p=0.001$. For this comparison, a random subset of $200$ subjects is used from each subject category. H and H* are two different random subset from the healthy subject pool.}
    \label{fig:stat_sig}
\vspace{-0.25in}
\end{figure}
\vspace{-0.1in}
\section{Results}

\subsection{Statistical Analysis}
The statistical significance obtained following the approach described in Section~\ref{sec:statistical_analysis} is shown in Figure~\ref{fig:stat_sig}. A higher value for $-\log_{10}(p)$, which is $>3$, indicates that the acoustic features compared between the two subject populations are statistically significantly different. For a fair comparison in terms of population size, a random subset of $200$ subjects is considered for each subject category. As expected, there is no significant
\begin{figure*}[t!]
    \centering
    \input{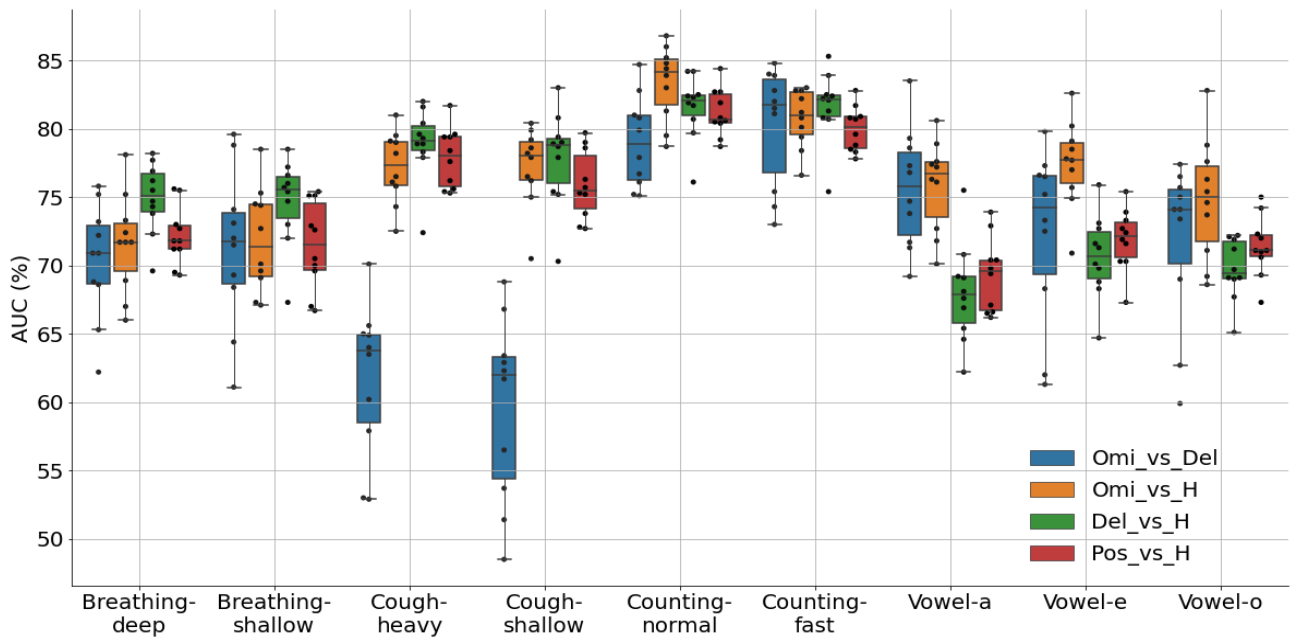}
    \vspace{-0.25in}
    \caption{10 fold validation AUCs for 9 categories of sounds and for the 4 classification tasks.}
    \label{fig:val_auc}
 \vspace{-0.05in}
\end{figure*}
\begin{table*}[]
\rowcolors{2}{white}{gray!10}
\begin{tabular}{p{1.2cm}p{1.2cm}p{1.2cm}p{1.2cm}p{1.2cm}p{1cm}p{1cm}p{0.8cm}p{0.8cm}p{0.8cm}p{2cm}}
\hline
% \rowcolor[HTML]{9B9B9B} 
\textbf{Task} & \bf Breathing-deep &\bf Breathing-shallow & \bf Cough-heavy & \bf Cough-shallow & \bf Counting-fast  & \bf Counting-normal & \bf Vowel-[\ae] & \bf Vowel-[e] & \bf Vowel-[o] & \bf Fusion (Sensitivity at 95\% Specificity)\\ \hline
    \textbf{Omi-Del}      &       79.9      &     73.4        & 65.2  &   65.7  & 81.4  &  79.8 &  73.0 & 75.3  &  77.9 &  89.0 (52.4)   \\ \hline
    \textbf{Del-H}       &      73.3       &     70.1        & 72.2  &  75.0   & 81.3   & 80.5  & 70.9  & 70.3  & 71.1 & 88.3 (68.6)\\ \hline
     \textbf{Omi-H}      &     80.2        &     76.9        & 81.2  &   83.7  &  84.0 & 82.1  & 75.8  & 79.5  & 74.9  & 93.1 (71.4) \\ \hline
     \textbf{Pos-H}       &      76.1       &    74.4         & 76.7  &  80.1   & 83.4  & 82.0  & 75.4  & 76.6  &  75.0 & 90.5 (71.4) \\ \hline
\end{tabular}
\caption{Depiction of AUC for different classification tasks on the held out test set.}
\label{test_auc_table}
\vspace{-0.25in} 
\end{table*}

% \begin{table*}[]
% \rowcolors{2}{white}{gray!10}
% \begin{tabular}{p{2.5cm}p{1cm}p{1cm}p{1cm}p{1cm}}
% \hline
% \rowcolor[HTML]{9B9B9B} 
% \textbf{Sound Category} & \textbf{O vs D} & \textbf{D vs H} & \textbf{O vs H} & \textbf{P vs H}\\ \hline
% breathing-deep          &             &             &   &          \\ \hline
% breathing-shallow       &            &            &  &          \\ \hline
% cough-heavy             &             &            &    &         \\ \hline
% Cough-shallow           &             &             &  &           \\ \hline
% counting-fast           &             &            &    &         \\ \hline
% counting-normal         &             &             &   &          \\ \hline
% vowel-a                 &             &            &   &          \\ \hline
% vowel-e                 &           &            &   &          \\ \hline
% vowel-o                 &             &            &    &         \\ \hline
% \end{tabular}
% \caption{Table of test AUCs}
% \label{val_auc_table}
% \end{table*}
\begin{figure}[t]
    \centering
    \includegraphics[width=4.5cm, height=4.1cm]{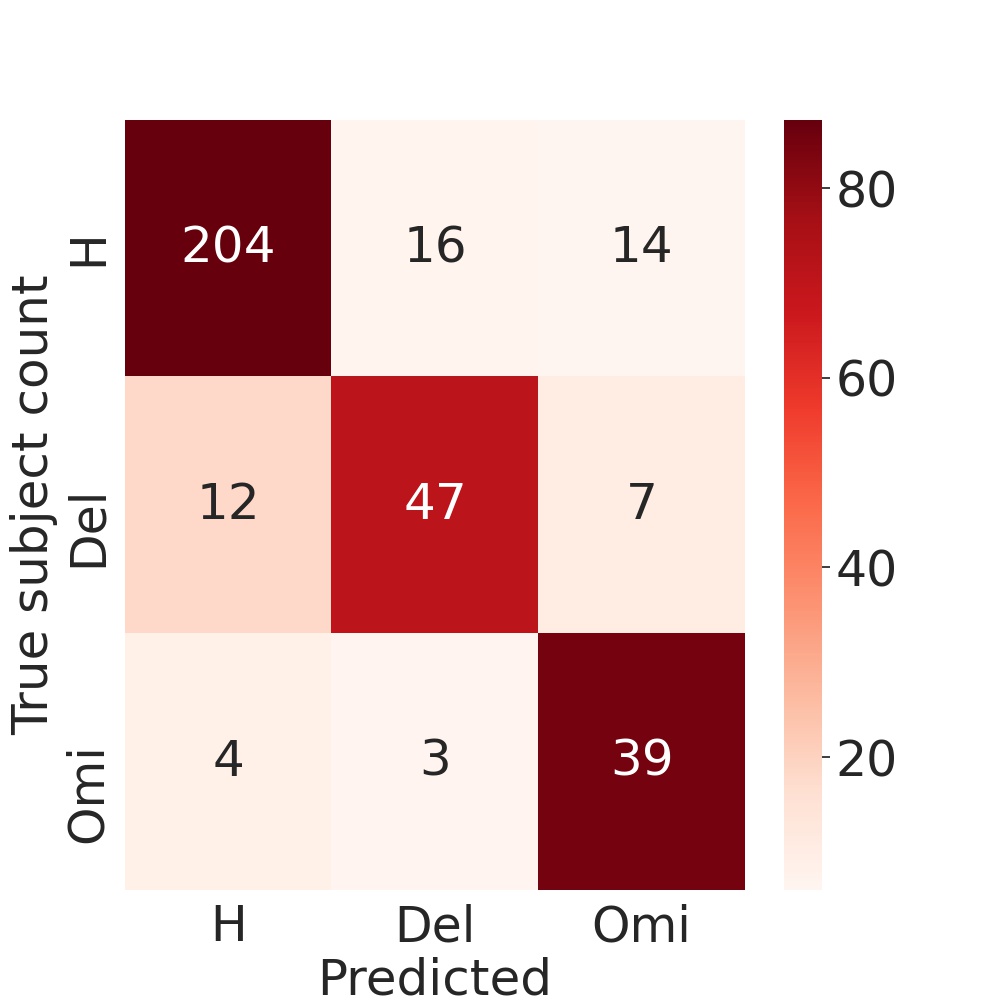}
    \caption{Confusion matrix for hierarchical $3$ class classification}
    \label{fig:confusion_matrix}
\vspace{-0.25in}
\end{figure}
difference between two different subsets of the healthy pool (denoted as $H$ and $H^{*}$). The significance increases for $Omi$-vs-$H$ and $Del$-vs-$H$ subject categories. This suggests that a binary classifier can be designed to separate healthy subjects from COVID-19 positive subjects. Interestingly, for a majority of sound modalities, there is also a significant difference between $Del$ and the $Omi$ subject categories, indicating that there are significant acoustic differences for the different variants of the SARS-CoV-2.
\subsection{Classification Results}
Four categories of two class classification problems are set up namely, $Omi$-vs-$Del$, $Omi$-vs-$H$, $Del$-vs-$H$ and $Pos$-vs-$H$ as described in section 3.4.
%For each task, first group is considered as positive and other group is considered as negative class. 
%Then positive and negative class samples are separately split for 10 different train and validation splits with 10 different random seeds.
The AUC distribution for the $9$ sound modalities and the $4$ classification tasks using the $10$ validation folds is shown in Figure \ref{fig:val_auc}. As seen in the figure, the cough sound modality does not provide a good separation in the $Omi$-vs-$Del$ classification. The speech counting is the best performing modality for $Omi$-vs-$Del$ classification task. The performance of the vowel sound based classifiers is inferior to the other modalities of breathing sounds and speech-counting modalities. 

Further, the test AUC values are computed, as given in Table \ref{test_auc_table}. Among all the sound modalities, the counting-fast is the category that provides more than $80$\% ROC-AUC for all the $4$ binary classifications tasks.
Further, a simple fusion of the acoustic scores from the different modalities (average of probability scores from each of the classifiers trained on the individual modalities), achieves an AUC of $93.1$\% for classifying $Omi$-vs-$H$ and 
$90.5$\% AUC for the $Pos$-vs-$H$ classification. 
A hierarchical classifier, designed using a serial combination of $Pos$-vs-$H$ followed by the $Omi$-vs-$Del$ classifier yields a diagonally dominant confusion matrix on the test data, as shown in Figure~\ref{fig:confusion_matrix}.  The confusion matrix shows that the proposed modeling and classification approaches allow the separation of healthy and positive samples, while further indicating the possibilty of variant idenfication within the positive samples. 
% \begin{figure}[t]
%     \centering
%     \includegraphics[width=8cm, height=6.25cm]{figures/confusion_matrix_3way.jpg}
%     \caption{Confusion matrix for direct three class classification}
%     \label{fig:stat_sig}
% \vspace{-0.25in}
% \end{figure}

% \begin{table}[]
% \begin{tabular}{|l|l|l|l|}
% \hline
% \rowcolor[HTML]{9B9B9B} 
% \textbf{Sound Category} & \textbf{O vs D} & \textbf{D vs H} & \textbf{O vs H} \\ \hline
% breathing-deep          & 75.0            & 75.8            & 76.2            \\ \hline
% breathing-shallow       & 57.1            & 72.5            & 73.4            \\ \hline
% cough-heavy             & 48.9            & 80.6            & 73.7            \\ \hline
% Cough-shallow           & 44.2            & 80.5            & 73.5            \\ \hline
% counting-fast           & 81.4            & 82              & 84.4            \\ \hline
% counting-normal         & 77.5            & 82.6            & 82.5            \\ \hline
% vowel-a                 & 58.2            & 70.5            & 73.8            \\ \hline
% vowel-e                 & 60.0            & 75.7            & 71.7            \\ \hline
% vowel-o                 & 69.8            & 74.4            & 71.5            \\ \hline
% \end{tabular}
% \caption{Table of validation AUCs}
% \label{val_auc_table}
% \end{table}

%\section{Discussion}
%- cough is not a good indicator for Omi vs Delta
%- detection above chance

% \subsection{Trustworthiness}

\section{Conclusion}
% We analyzed the acoustic features extracted from nine different categories, collected from healthy and SARS-CoV-2 infected subjects in India. 
Analyzing the acoustic features of healthy and SARS-CoV-2 infected subjects in India,
we find a statistically significant difference when comparing a pair of subject categories, namely, healthy versus COVID-19 positive (both for delta and omicron variants). Further, there is also a significant difference in acoustic features for audio collected from COVID-19 positive subjects with delta variant versus subjects with omicron variant. The binary classifiers designed to distinguish healthy from COVID-19 positive subjects  performed significantly above chance across the different pairs of subject categories. 
Our findings suggest that the impact on the respiratory health of the COVID-19 subject with the omicron variant is different from that of the patients infected by the delta variant. While this study is not based on genomic confirmation of the SARS-CoV-2 variants on the subject pools, we still hypothesize that the insights developed in this paper encourage a more detailed study to allow the identification of SARS-Cov-2 variants from audio samples.

\section{Acknowledgement}
\noindent  The authors thank the Department of Science and Technology, Govt. of India, for funding the Coswara Project through the RAKSHAK program. The authors express their gratitude to - all the individuals who voluntarily contributed to the dataset, Anand Mohan for designing the web-based audio data collection platform, and Aditya Jana, Drishti Gupta, Priyanka Ramesh and Riya Bhattacharya for coordinating the data collection.
\bibliographystyle{IEEEtran}

\bibliography{mybib}

\begin{thebibliography}{10}
\providecommand{\url}[1]{#1}
\csname url@samestyle\endcsname
\providecommand{\newblock}{\relax}
\providecommand{\bibinfo}[2]{#2}
\providecommand{\BIBentrySTDinterwordspacing}{\spaceskip=0pt\relax}
\providecommand{\BIBentryALTinterwordstretchfactor}{4}
\providecommand{\BIBentryALTinterwordspacing}{\spaceskip=\fontdimen2\font plus
\BIBentryALTinterwordstretchfactor\fontdimen3\font minus
  \fontdimen4\font\relax}
\providecommand{\BIBforeignlanguage}[2]{{%
\expandafter\ifx\csname l@#1\endcsname\relax
\typeout{** WARNING: IEEEtran.bst: No hyphenation pattern has been}%
\typeout{** loaded for the language `#1'. Using the pattern for}%
\typeout{** the default language instead.}%
\else
\language=\csname l@#1\endcsname
\fi
#2}}
\providecommand{\BIBdecl}{\relax}
\BIBdecl

\bibitem{pramesh2021choosing}
C.~Pramesh, G.~R. Babu, J.~Basu, I.~Bhushan, C.~M. Booth, G.~Chinnaswamy,
  R.~Guleria, S.~Kalantri, G.~Kang, P.~Mohan \emph{et~al.}, ``Choosing wisely
  for {COVID-19}: ten evidence-based recommendations for patients and
  physicians,'' \emph{Nature Medicine}, vol.~27, no.~8, pp. 1324--1327, 2021.

\bibitem{he2021sars}
X.~He, W.~Hong, X.~Pan, G.~Lu, and X.~Wei, ``Sars-cov-2 omicron variant:
  characteristics and prevention,'' \emph{MedComm}, 2021.

\bibitem{mohapatra2022twin}
R.~K. Mohapatra, R.~Tiwari, A.~K. Sarangi, S.~K. Sharma, R.~Khandia,
  G.~Saikumar, and K.~Dhama, ``Twin combination of omicron and delta variant
  triggering a tsunami wave of ever high surges in covid-19 cases: a
  challenging global threat with a special focus on {I}ndian sub-continent,''
  \emph{Journal of medical virology}, 2022.

\bibitem{gopal2022analysis}
R.~Gopal, V.~Chandrasekar, and M.~Lakshmanan, ``Analysis of the second wave of
  {COVID-19 in India based on SEIR} model,'' \emph{The European Physical
  Journal Special Topics}, pp. 1--8, 2022.

\bibitem{ranjan2022omicron}
R.~Ranjan, ``Omicron impact in india: An early analysis of the ongoing
  {COVID-19} third wave,'' \emph{medRxiv}, 2022.

\bibitem{hui2022sars}
K.~P. Hui, J.~C. Ho, M.-c. Cheung, K.-c. Ng, R.~H. Ching, K.-l. Lai, T.~T. Kam,
  H.~Gu, K.-Y. Sit, M.~K. Hsin \emph{et~al.}, ``Sars-cov-2 omicron variant
  replication in human bronchus and lung ex vivo,'' \emph{Nature}, pp. 1--5,
  2022.

\bibitem{sharma2020coswara}
N.~Sharma, P.~Krishnan, R.~Kumar, S.~Ramoji, S.~R. Chetupalli, R.~Nirmala,
  P.~K. Ghosh, and S.~Ganapathy, ``Coswara -- a database of breathing, cough,
  and voice sounds for {{COVID}-19} diagnosis,'' in \emph{Proc. Interspeech},
  2020, pp. 4811--4815.

\bibitem{tinkelman1991analysis}
D.~G. Tinkelman, C.~Lutz, and B.~Conner, ``Analysis of breath sounds in normal
  and asthmatic children and adults using computer digitized airway
  phonopneumography (cdap),'' \emph{Respiratory medicine}, vol.~85, no.~2, pp.
  125--131, 1991.

\bibitem{al2012asthma}
M.~Al-khassaweneh, S.~B. Mustafa, and F.~Abu-Ekteish, ``Asthma attack
  monitoring and diagnosis: A proposed system,'' in \emph{2012 IEEE-EMBS
  Conference on Biomedical Engineering and Sciences}.\hskip 1em plus 0.5em
  minus 0.4em\relax IEEE, 2012, pp. 763--767.

\bibitem{nathan2019assessment}
V.~Nathan, K.~Vatanparvar, M.~M. Rahman, E.~Nemati, and J.~Kuang, ``Assessment
  of chronic pulmonary disease patients using biomarkers from natural speech
  recorded by mobile devices,'' in \emph{2019 IEEE 16th International
  Conference on Wearable and Implantable Body Sensor Networks (BSN)}.\hskip 1em
  plus 0.5em minus 0.4em\relax IEEE, 2019, pp. 1--4.

\bibitem{mann1947test}
H.~B. Mann and D.~R. Whitney, ``On a test of whether one of two random
  variables is stochastically larger than the other,'' \emph{The Annals of
  mathematical statistics}, pp. 50--60, 1947.

\bibitem{Wilson1195}
D.~J. Wilson, ``The harmonic mean p-value for combining dependent tests,''
  \emph{Proceedings of the National Academy of Sciences}, vol. 116, no.~4, pp.
  1195--1200, 2019.

\bibitem{sharma2021second}
N.~K. Sharma, S.~R. Chetupalli, D.~Bhattacharya, D.~Dutta, P.~Mote, and
  S.~Ganapathy, ``The {S}econd {D}icova {C}hallenge: {D}ataset and performance
  analysis for {COVID-19} diagnosis using acoustics,'' \emph{accepted in IEEE
  ICASSP}, 2022.

\bibitem{FAWCETT2006861}
T.~Fawcett, ``An introduction to {ROC} analysis,'' \emph{Pattern Recognition
  Letters}, vol.~27, no.~8, pp. 861--874, 2006, rOC Analysis in Pattern
  Recognition.

\bibitem{scikit-learn}
F.~Pedregosa, G.~Varoquaux, A.~Gramfort, V.~Michel, B.~Thirion, O.~Grisel,
  M.~Blondel, P.~Prettenhofer, R.~Weiss, V.~Dubourg, J.~Vanderplas, A.~Passos,
  D.~Cournapeau, M.~Brucher, M.~Perrot, and E.~Duchesnay, ``Scikit-learn:
  Machine learning in {P}ython,'' \emph{Journal of Machine Learning Research},
  vol.~12, pp. 2825--2830, 2011.

\end{thebibliography}

% \begin{thebibliography}{9}
% \bibitem[1]{Davis80-COP}
%   S.\ B.\ Davis and P.\ Mermelstein,
%   ``Comparison of parametric representation for monosyllabic word recognition in continuously spoken sentences,''
%   \textit{IEEE Transactions on Acoustics, Speech and Signal Processing}, vol.~28, no.~4, pp.~357--366, 1980.
% \bibitem[2]{Rabiner89-ATO}
%   L.\ R.\ Rabiner,
%   ``A tutorial on hidden Markov models and selected applications in speech recognition,''
%   \textit{Proceedings of the IEEE}, vol.~77, no.~2, pp.~257-286, 1989.
% \bibitem[3]{Hastie09-TEO}
%   T.\ Hastie, R.\ Tibshirani, and J.\ Friedman,
%   \textit{The Elements of Statistical Learning -- Data Mining, Inference, and Prediction}.
%   New York: Springer, 2009.
% \bibitem[4]{YourName17-XXX}
%   F.\ Lastname1, F.\ Lastname2, and F.\ Lastname3,
%   ``Title of your INTERSPEECH 2022 publication,''
%   in \textit{Interspeech 2022 -- 23\textsuperscript{rd} Annual Conference of the International Speech Communication Association, September 18-22, Incheon, Korea, Proceedings, Proceedings}, 2022, pp.~100--104.
% \end{thebibliography}

\end{document}